\documentclass[aps,prl,twocolumn,showkeys,floatfix,superscriptaddress]{revtex4}

\usepackage{graphicx}
\usepackage{dcolumn}
\usepackage{bm}

\newcommand{\erf}{\mbox{erf}}

\newcommand{\pdiffl}[2]{\frac{\partial #1}{\partial #2}}

\newcommand{\dfrac}[2]{\displaystyle\frac{#1}{#2}}

\newcommand{\kB}{k_{\text{B}}}
\newcommand{\rWS}{r_{\text{WS}}}
\newcommand{\thetaD}{\theta_{\text{D}}}
\newcommand{\thetaE}{\theta_{\text{E}}}
\newcommand{\xD}{x_{\text{D}}}
\newcommand{\xE}{x_{\text{E}}}

\begin{document}


\title{High temperature ion-thermal behavior from average-atom calculations}

\date{November 1, 2018; revisions to March 13, 2019
  -- LLNL-JRNL-768634}

\author{Damian C. Swift}
\email{dswift@llnl.gov}
\affiliation{%
   Lawrence Livermore National Laboratory,
   7000 East Avenue, Livermore, California 94551, USA
}
\author{Mandy Bethkenhagen\footnote{%
   Present affiliation: Universit\"at Rostock, 18051 Rostock, Germany
}}
\affiliation{%
   Lawrence Livermore National Laboratory,
   7000 East Avenue, Livermore, California 94551, USA
}
\author{Alfredo A. Correa}
\affiliation{%
   Lawrence Livermore National Laboratory,
   7000 East Avenue, Livermore, California 94551, USA
}
\author{Thomas Lockard}
\affiliation{%
   Lawrence Livermore National Laboratory,
   7000 East Avenue, Livermore, California 94551, USA
}
\author{Sebastien Hamel}
\affiliation{%
   Lawrence Livermore National Laboratory,
   7000 East Avenue, Livermore, California 94551, USA
}
\author{Lorin X. Benedict}
\affiliation{%
   Lawrence Livermore National Laboratory,
   7000 East Avenue, Livermore, California 94551, USA
}
\author{Philip A. Sterne}
\affiliation{%
   Lawrence Livermore National Laboratory,
   7000 East Avenue, Livermore, California 94551, USA
}
\author{Bard I. Bennett}
\affiliation{%
   Los Alamos National Laboratory,
   PO Box 1663, Los Alamos, New Mexico 87545, USA
}

\begin{abstract}
Atom-in-jellium calculations of the Einstein frequency were used
to calculate the mean displacement of an ion over a wide range of compression
and temperature.
Expressed as a fraction of the Wigner-Seitz radius, the displacement
is a measure of the asymptotic freedom of the ion at high temperature,
and thus of the change in heat capacity from 6 to 3 quadratic degrees of
freedom per atom.
A functional form for free energy was proposed based on the Maxwell-Boltzmann
distribution as a correction to the Debye free energy, 
with a single free parameter representing the effective density of potential
modes to be saturated.
This parameter was investigated using molecular dynamics simulations,
and found to be $\sim$0.2 per atom. 
In this way, the ion-thermal contribution can be
calculated for a wide-range equation of state (EOS) without requiring a large number
of molecular dynamics simulations.
Example calculations were performed for carbon,
including the sensitivity of key EOS loci to ionic freedom.
\end{abstract}

\keywords{equation of state}

\maketitle

\section{Introduction}
Accurate equations of state (EOS) are essential to understand stellar and 
planetary formation and evolution, astrophysical impacts, and engineering
challenges such as the development of thermonuclear energy sources. However,
the behavior of the ionic heat capacity of condensed matter at temperatures
between melting and the formation of an ideal plasma is poorly understood,
limiting the insight and accuracy of theoretical EOS.
Wide-ranging EOS \cite{sesleos}
are almost invariably constructed using empirical models
originally derived as approximate representations of observations
of the variation of the heat capacity in the liquid 
of metals of low melting point at one atmosphere \cite{Grover1971},
and assumed to apply at arbitrary compression
\cite{cowan,Johnson1991}.
Experiments to test or improve on this assumption are challenging:
where even attempted, uncertainties on measurements of the temperature 
of warm dense matter are typically greater than 10\%\ 
(see for instance \cite{Kritcher2011,Saunders2016}),
which is not adequate to distinguish
the details of the ion-thermal heat capacity.

The most rigorous theoretical techniques applicable are path-integral
Monte-Carlo (PIMC) and quantum molecular dynamics (QMD), in which the 
kinetic motion of an ensemble of atoms is simulated, where the distribution
of the electrons is found with respect to the changing location of the ions
using quantum mechanics \cite{pimc,qmd}.
The energy of the ensemble is determined from an average over a sufficient
time interval, and the heat capacity can be found from the variation
of energy with temperature.
This procedure is computationally expensive, requiring $o(10^{16})$ or more
floating-point operations per state to determine the ionic heat capacity
using QMD, 
equivalent to thousands of CPU-hours per state; PIMC requires roughly
an order of magitude more.
It is typically deemed impractical to perform these simulations for matter
around or below ambient density and above a few tens of electron volts 
using QMD, limiting the regions of state space over which the EOS
can be calibrated in this way.

Recent PIMC and QMD results have indicated that the simpler approach of
calculating the electron states for a single atom in a spherical cavity
within a uniform charge density of ions and electrons, representing the
surrounding atoms, reproduces the electronic component of 
their more rigorous EOS \cite{Benedict2014,Driver2017}.
This atom-in-jellium approach \cite{Liberman1979} has been used previously 
to predict the electron-thermal energy of matter at high temperatures and
compressions, as an advance over Thomas-Fermi and related approaches \cite{tf}.
A development of atom-in-jellium was used to estimate ion-thermal properties
using the Debye model \cite{Liberman1990},
and we have found that it can be used to construct the complete EOS
\cite{Swift2018}.
However, the model as originally implemented did not account for the decrease in
ionic heat capacity at high temperatures as the ions cease to be caged
between their neighbors, losing the contribution from potential energy.

In the work reported here, we extend the atom-in-jellium ion-thermal model 
developed previously to estimate the asymptotic freedom of ions at high
temperature, and hence predict the form of the heat capacity of matter
in the fluid-plasma regime.
Because of the efficiency of average-atom calculations, this approach should 
for the first time
enable EOS to be constructed with consistently 
accurate electronic contributions and
the correct ion-thermal behavior over a wide range of states from
expanded to compressed matter and over the full range of temperatures relevant
to stellar and planetary interiors, and thermonuclear fusion science.

\section{Previous ion-thermal models}
As condensed matter is heated from absolute zero, the heat capacity of the
ions rises from zero as vibrational modes are excited.
If all modes are excited before any dissociation occurs, the ionic heat capacity
for quadratic degrees of freesom
reaches $3 \kB$ per atom, where $\kB$ is the Boltzmann constant, representing
3 kinetic and 3 potential modes \cite{Waldram1985}.
The attractive potential between atoms has a finite depth, and once an ion
has more energy, it behaves as a free particle with only kinetic modes 
available to it, and thus contributes $3 \kB/2$ to the heat capacity.

The instantaneous separation and velocity of the ions are described by a 
distribution, and the most appropriate description of the ion energies is also
by a distribution. Even at low temperatures some ions are free, and even at
high temperatures some ions are bound.
The detailed distribution of ion energies depends on the shape of the 
interatomic potential as well as the temperature, which complicates the
analysis.

EOS have been constructed using very simple ion-thermal models,
such as a constant specific heat capacity which may be the value at STP,
$3 \kB$ per atom, or $3 \kB/2$ per atom \cite{simplecv}.
A generalization has been to use the Debye model \cite{debye},
which assumes a simple form for the phonon density of states (PDOS),
proportional to $\hbar\omega^2$ for phonon frequencies 
$0\le\omega\le\kB\thetaD$, and 0 for higher $\omega$.
This model captures the rise of ion-thermal heat capacity 
from zero to $3 \kB$ per atom as modes are excited, but
does not capture the detailed heat capacity arising from the actual PDOS.
EOS can be constructed using more accurate PDOS \cite{Swift2001}, 
though, in practice, because integrations are performed over the PDOS,
the EOS is not sensitive to the full detail of the spectrum,
and high-fidelity EOS have been constructed 
using a combination of a few Debye frequencies instead \cite{multidebye}.
More importantly for the present study, the Debye model
ignores the asymptotic freedom of the ions at high temperature.

In the ion thermal model developed for use with atom-in-jellium calculations
\cite{Liberman1990},
perturbation theory was used to calculate the Hellmann-Feynman force on the
ion when displaced from the center of the cavity in the jellium.
Given the force constant $k=-\partial f/\partial r$, the Einstein vibration
frequency $\nu_e=\sqrt{k/m_a}$ was determined, where $m_a$ is the atomic mass,
and hence the Einstein temperature $\thetaE=h\nu_e/\kB$.
The Debye temperature $\thetaD$ was inferred from $\thetaE$, by
equating either the ion-thermal energy $e_i$
or the mean square displacement $\bar{u^2}$, giving slightly different results.
These calculations are, respectively,
\begin{equation}
e_i\frac{m_a}{3\kB T} = 
\xE\left[\frac 1{\exp(\xE)+1}+\frac 12\right]
 = D_3(\xD)+\frac 38\xD
\end{equation}
and
\begin{equation}
\bar{u^2} \frac{m_a}{3\hbar^2} =
\dfrac 1{\thetaE}\left[\dfrac 1{\exp(\xE)-1}+\frac 12\right]
 =
\frac 1{\thetaD}\left[\frac{D_1(\xD)}{\xD}+\frac 14\right]
\end{equation}
where $\xE=\thetaE/T$, $\xD=\thetaD/T$,
and $D_i$ is the Debye integral
\begin{equation}
D_i(x)\equiv\dfrac i{x^i}\int_0^x\dfrac{x^i\,dx}{e^x-1}.
\end{equation}
Below, we use the displacement calculation, for consistency.
The ion-thermal free energy was then calculated from
\begin{equation}
f_i = \kB T\left[3\ln\left(1-e^{-\thetaD/T}\right)+\frac{9\thetaD}{8T}
-D_3(\thetaD/T)\right]
\end{equation}
where $\frac 98 \kB\thetaD$ is the zero-point energy.
Unusually, $\thetaD$ is a function of temperature as well as compression,
effectively because changes in ionization can result in a change to the
stiffness and hence the vibration frequency. 

An approach used widely in constructing EOS for fluid-plasma applications 
is the Cowan model \cite{cowan},
in which the heat capacity is assumed to vary as 
\begin{equation}
c_v=\frac 32 \kB\left\{1+\mbox{min}\left[1,\left(\frac T{T_m(\rho)}\right)^{-1/3}\right]\right\}
\end{equation}
where $T_m(\rho)$ is the melting temperature as a function of mass density 
$\rho$.
The effect of the term in brackets is for the potential modes to fall as 
$(T/T_m)^{-1/3}$ for $T > T_m$, and for $c_v$ to be held at 
$3 \kB$ for $T\le T_m$.
Variants of the Cowan model have been used with more general
dependence on $T/T_m$, also treating the latent heat of melting in
{\it ad hoc} fashion as an increased ionic heat capacity over a finite
temperature range \cite{Johnson1991}.

QMD studies of carbon \cite{Benedict2014}
found that the heat capacity 
dropped more abruptly than in the Cowan model, and were represented better
as a free energy of the form
\begin{equation}
f_i=f_b-\kB T\ln\left[\erf\left(\sqrt{\dfrac{T_r}T}\right)-\sqrt{\frac{4 T_r}{\pi T}}e^{-T_r/T}\right]
\end{equation}
where $f_b$ is the free energy of bound ions and
$T_r(\rho)$ is a reference temperature curve determined from QMD.
This approach appears to be as accurate as QMD can achieve,
but is limited by the restricted range of states accessible in practice to QMD.

\section{Asymptotic freedom of ions in jellium}
In the method developed for calculating the vibrations of ions in jellium
\cite{Liberman1990},
the Debye temperature $\thetaD$ was inferred from the Einstein temperature
$\thetaE$ by equating the energy or displacement, as discussed above.
The mean displacement $u$ can be used as a measure of ionic freedom:
when it exceeds the Wigner-Seitz radius,
$\rWS=\left({3 m_a}/{4 \pi \rho}\right)^{1/3}$, 
the ion is effectively free.

\begin{figure}
\begin{center}\includegraphics[scale=0.72]{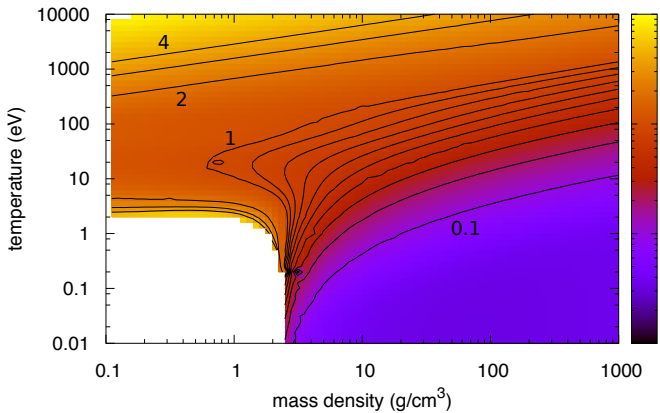}\end{center}
\caption{Mean fractional displacement calculated for carbon.
   Contours shown are from 0.1 to 1.0 at intervals of 0.1, and then 2.0 to 4.0
   at intervals of 1.0.
   A fractional displacement of 1 would correspond to ionic freedom,
   ignoring the velocity distribution of the ions.}
\label{fig:disp}
\end{figure}

The {\sc Inferno} program was modified to calculate and output the mean
fractional displacement, $u/\rWS(\rho,T)$ (Fig.~\ref{fig:disp}).

An average atom would be bound for $u < \rWS$, with ionic heat capacity
$3 \kB$, and then free for $u\ge \rWS$, with heat capacity $3\kB/2$.
Given an energy distribution for the atoms, the change in heat capacity can
be represented more accurately as a continuous variation with temperature.
An accurate distribution could be calculated from the set of available
ion-thermal energy levels populated using Boltzmann factors,
but the average-atom-in-jellium method gives only a rough approximation to the
states and hence to the energy levels.
A simpler estimate can be made using the Maxwell-Boltzmann distribution
for the velocity $v$ of free ions,
\begin{equation}
f(v) = \left(\frac{m_a}{2\pi \kB T}\right)^{3/2} 4\pi v^2 e^{-m_av^2/2\kB T}
\end{equation}
\cite{Waldram1985}.
The cumulative probability distribution, giving the fraction of ions whose
velocity is less than $v$, is
\begin{equation}
Pr(< v) = \erf\left(\sqrt{\frac{m_av^2}{4\kB T}}\right)-
   \sqrt{\frac{2m_av^2}{\pi \kB T}}e^{-m_av^2/2\kB T}.
\end{equation}
The critical displacement $u=\rWS$ can be equated to a critical velocity $v$, 
and hence to a characteristic binding temperature,
\begin{equation}
\frac 12 m_av^2\simeq \frac N2 \kB T_b,
\end{equation}
where $N$ is the number of modes that must be saturated for the ion 
to become free, i.e.
\begin{equation}
T_b\equiv T : u = \rWS.
\end{equation}
We have not so far found an integral of the Maxwell-Boltzmann
derived heat capacity to represent the free energy in closed form,
but we can generalize the similar functional form used previously 
\cite{Benedict2014} and derived from the partition function of a particle
in a harmonic potential of finite volume \cite{Correa2018},
which we have shown exhibits the desired temperature
dependence for the heat capacity, giving
a modification to the Debye free energy 
(or any other free energy $f_b$ representing bound ions)
so that the heat capacity falls at high temperature, 
\begin{equation}
f_i = f_b -
 \kB T \ln\left[\erf\left(\sqrt{\frac{NT_b}T}\right)
 - \sqrt{\frac{4NT_b}{\pi T}} e^{-NT_b/T}\right].
\label{eq:cowanx}
\end{equation}
This approach is similar, except that $T/T_b(\rho,T)$ is determined directly
from the average-atom calculations, and the mode number $N$ is likely to be
a constant $o(1)$ with $\rho$, $T$, and atom type.
In contrast, the approach used in the previous study \cite{Benedict2014}
requires QMD simulations to be performed for at least a few temperatures
and over the full range of densities of interest, for each substance.

\section{Quantum molecular dynamics simulations}
QMD simulations were performed to predict the variation of
ionic heat capacity $c_{vi}$ directly.
Such simulations treat the motion of the ions as classical,
with 3 kinetic modes.
The total potential energy is calculated from the electron states with respect
to the instantaneous configuration of the ions at the temperature of interest.
The potential contribution to the $c_{vi}$ must thus be inferred 
from the total $c_v$, 
by calculating and subtracting the electronic heat capacity $c_{ve}$
\cite{Whitley2015}.
Along each isochore, 
$c_{vi}$ was found to fall just as $c_{ve}$ started to rise,
so the deduced variation in $c_{vi}(T)$ was sensitive to the treatment of 
$c_{ve}$.
In addition, the drop in $c_{vi}$ started to become appreciable 
a little above the melt locus,
so its precise variation depended on discriminating the latent heat of melting,
which may be spread out in temperature in a relatively small ensemble of atoms.

\begin{figure}
\begin{center}\includegraphics[scale=0.72]{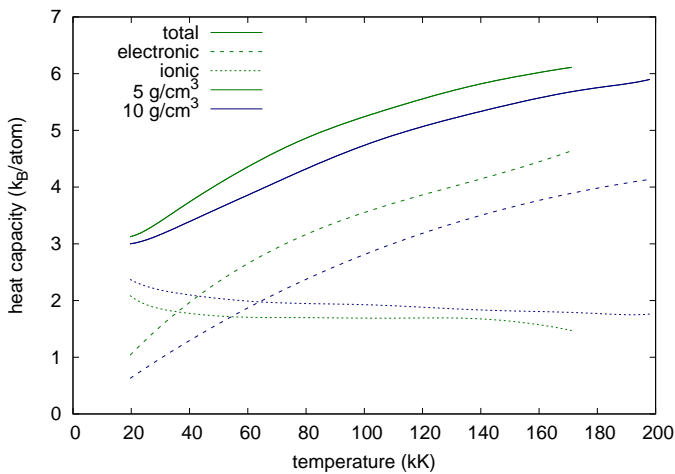}\end{center}
\caption{Variation of heat capacities along sample isochores.
   Fully activated vibrational modes give $3\kB$/atom, and
   unbound kinetic modes $\frac 32\kB$/atom.}
\label{fig:cvcmp}
\end{figure}

QMD simulations were performed using the electronic structure
program {\sc Vasp} \cite{vasp}.
The projector augmented wave method \cite{paw} was used,
with carbon ions represented with a pseudopotential
subsuming the inner two electrons.
Electron wavefunctions were represented with a plane wave basis set 
cut off at 1000\,eV,
at the Baldereschi mean-value point in reciprocal space \cite{Baldereschi1972}.
Density functional theory in the local density approximation
\cite{Hohenberg1964,Kohn1965,Perdew1992,Perdew1994} was used
for the exchange-correlation energy;
some states were recalculated with the Perdew-Burke-Ernzerhof functional
\cite{Perdew1996} for comparison.
The simulations were in the NVT ensemble, using the Nos\'e-Hoover thermostat
\cite{Hoover1996}, with periodic boundary conditions.
For each state of density and temperature, the motion of 64 atoms 
was integrated for 20000-50000 steps of 0.05-0.5\,fs.
Convergence with respect to plane-wave cutoff energy was tested 
at 2000\,K and 5\,g/cm$^3$ with calculations up to 3000\,eV,
and found to be converged to $<$0.06\%\ in pressure and $<$4\,meV/atom.
Overall, given the finite cell size and use of the single mean-value
$k$-point, the pressure is likely to be converged to $<$2\%\ 
and the energy to $<$10\,meV/atom.
Simulations were performed along isochores at 5 and 10\,g/cm$^3$.
Each simulation yielded a value for the total energy $e$ and the electronic
entropy $s_e$, the latter from the population of electron states.
Using
\begin{equation}
c_v \equiv \left.\pdiffl eT\right|_v = T\left.\pdiffl sT\right|_v,
\end{equation}
the total heat capacity $c_v$ was deduced from $e(T)$,
and the electronic heat capacity $c_{ve}$ from $s_e(T)$, by fitting
polynomials to sections of each isotherm, and differentiating.
Given the numerical uncertainties in convergence, differentiation, and
subtraction, the uncertainty in ionic heat capacity was around 0.25\,$\kB$/atom.
(Fig.~\ref{fig:cvcmp}.)

\begin{figure}
\begin{center}\includegraphics[scale=0.72]{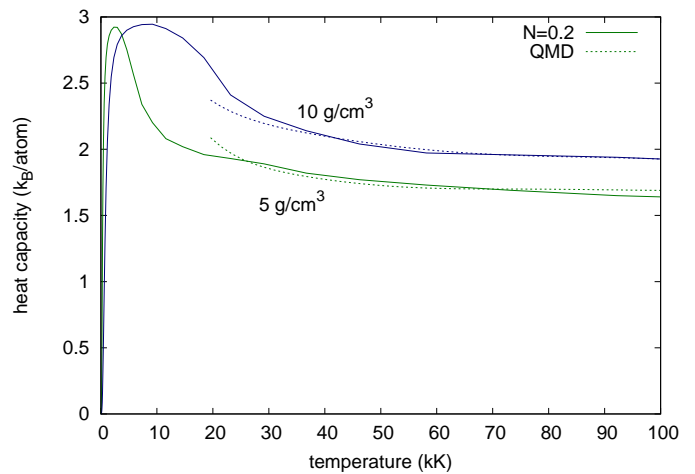}\end{center}
\caption{Variation of atom-in-jellium ion-thermal heat capacity 
   along sample isochores.}
\label{fig:cvfit}
\end{figure}

To interpret the QMD results,
the temperature along each isochore was expressed as 
$u^2/\rWS^2(\rho,T)\equiv T/T_b$ from the atom-in-jellium calculations.
The best fit of the hypothesized free energy function, Eq.~\ref{eq:cowanx},
was found for $N=0.2$
(Fig.~\ref{fig:cvfit}.)

\section{Equation of state for carbon}

Several wide-range EOS have been constructed for C, mostly using Thomas-Fermi
theory for high compressions and temperatures.
We compare against one such EOS employing a Cowan-like ion-thermal model,
{\sc sesame} EOS 7834 \cite{Crockett2006}, 
which was constrained by QMD calculations and Hugoniot measurements
up to shock melting, but does not include the melting transition explicitly.
In contrast, the 5-phase EOS \cite{Benedict2014} used 
atom-in-jellium calculations for the electron-thermal contribution only,
and was constructed to include an explicit melting transition.

\begin{figure}
\begin{center}\includegraphics[scale=0.72]{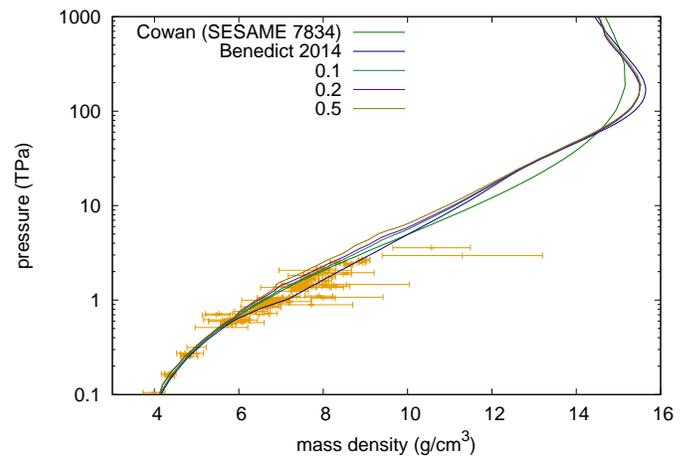}\end{center}
\caption{Principal shock Hugoniot for carbon calculated using different
   EOS, including atom-in-jellium with different values of the parameter
   $N$ in Eq.\ref{eq:cowanx},
   and compared with experimental data \cite{shockexpts}.}
\label{fig:hugdp}
\end{figure}

\begin{figure}
\begin{center}\includegraphics[scale=0.72]{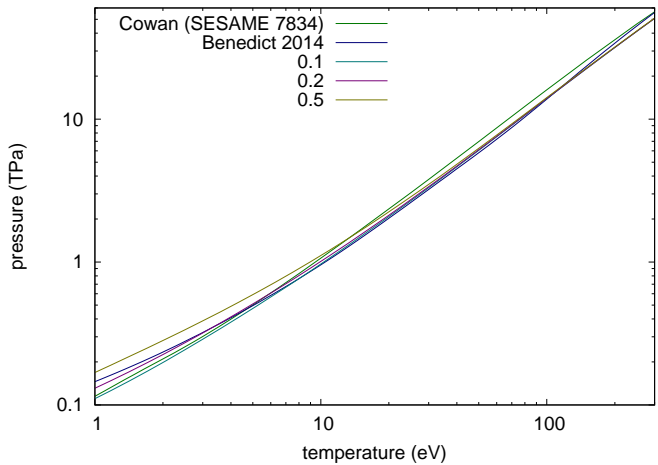}\end{center}
\caption{Ambient isochore for carbon from different EOS,
   including atom-in-jellium with different values of the parameter
   $N$ in Eq.\ref{eq:cowanx}.}
\label{fig:isochtp}
\end{figure}

The principal shock Hugoniot deduced by
numerical solution \cite{Swift2008,Swift2018a} for each EOS.
The treatment of ionic freedom caused a variation of up to 20\%\ in pressure
between 6 and 12\,g/cm$^3$ (1 and 10\,TPa), large enough to investigate
experimentally.
(Figs.~\ref{fig:hugdp} and \ref{fig:isochtp}.)

\section{Conclusions}
Atom-in-jellium calculations of thermal vibrations of the ion were used to
construct a model of the reduction in ionic heat capacity at high temperatures
as the ions become free.
The model has a single free parameter, equivalent to the effective number of
potential modes that must be saturated before the high-energy tail of the
Maxwell-Boltzmann distribution describes free particles.
This parameter was investigated using molecular dynamics simulations
and found to be $0.2\pm 0.03$, with a weak dependence on compression.

Carbon was chosen as a prototype material as its atomic number is low,
emphasizing changes in the ionic heat capacity compared with the electrons,
and it is widely used as a sample and ablator in high pressure experiments.
The principal shock Hugoniot and ambient isochore showed a modest sensitivity
to the treatment of ionic heat capacity, which may be experimentally detectable.

Atom-in-jellium calculations of electronic states 
were previously shown to be as accurate
for warm dense matter as the most rigorous approaches of
path-integral Monte Carlo and quantum molecular dynamics.
The work reported here makes
the atom-in-jellium based treatment of ion-thermal energies
as complete as these multi-atom calculations,
and means it is possible for the first time
to calculate all contributions to the equation of state
self-consistently,
and efficiently enough to construct an entire wide-range equation of state
for an element in a few CPU-hours at most.

\section*{Acknowledgments}

This work was performed under the auspices of
the U.S. Department of Energy under contract DE-AC52-07NA27344.


\begin{thebibliography}{10}
\bibitem{sesleos}{Prominent examples are the Los Alamos and Lawrence Livermore
      National Laboratories' EOS libraries:
   S.P.~Lyon and J.D.~Johnson, Los Alamos National Laboratory
      report LA-UR-92-3407 (1992);
   R.M.~More, K.H.~Warren, D.A.~Young and G.B.~Zimmerman,
      Phys. Fluids {\bf 31}, 3059 (1988);
   D.A.~Young and E.M.~Corey, J.~Appl. Phys. {\bf 78}, 3748 (1995).}
\bibitem{Grover1971}{R.~Grover, J.~Chem. Phys. {\bf 55}, 3435 (1971).}
\bibitem{cowan}{C.W.~Cranfill and R.~More, Los Alamos Scientific Laboratory
   report LA-7313-MS (1978);
   R.M.~More, K.H.~Warren, D.A.~Young, and G.B.~Zimmerman, Phys. Fluids {\bf 31}, 3059 (1988).}
\bibitem{Johnson1991}{J.D.~Johnson, High Press. Res. {\bf 6} (5), 277-285 (1991).}
\bibitem{Correa2018}{A.A.~Correa, L.X.~Benedict, M.A.~Morales, P.A.~Sterne,
   J.~I.~Castor, and E.~Schwegler, {\tt arXiv:1806.01346} (2018).}
\bibitem{Kritcher2011}{A.L.~Kritcher, T.~D\"oppner, C.~Fortmann, T.~Ma, O.L.~Landen, R.~Wallace, and S.H.~Glenzer,
   Phys. Rev. Lett. {\bf 107}, 015002 (2011).}
\bibitem{Saunders2016}{A.M.~Saunders, A.~Jenei, T.~D\"oppner, R.W.~Falcone,
   D.~Kraus, A.~Kritcher, O.L.~Landen, J.~Nilsen, and D.~Swift,
   Rev. Sci. Instrum. {\bf 87}, 11E724 (2016).}
\bibitem{pimc}{E.L.~Pollock and D.M.~Ceperley, Phys. Rev.~B {\bf 30}, 2555 (1984).}
\bibitem{qmd}{For example,
   L.~Collins, I.~Kwon, J.~Kress, N.~Troullier, and D.~Lynch,
   Phys. Rev.~E {\bf 52}, 6202 (1995).
   }
\bibitem{Benedict2014}{L.X.~Benedict, K.P.~Driver, S.~Hamel, B.~Militzer, T.~Qi, A.A.~Correa, A.~Saul, and E.~Schwegler,
   Phys. Rev. B {\bf 89}, 224109 (2014).
   Note:
   in Eq.~11 of this reference, the exponent is positive,
   which gives a heat capacity of $2.5 \kB$ as $T\rightarrow 0$.
   The negative exponent gives $3 \kB$, as desired.
}
\bibitem{Driver2017}{K.P.~Driver and B.~Militzer, Phys. Rev. E {\bf 95}, 043205 (2017).}
\bibitem{Liberman1979}{D.A.~Liberman, Phys. Rev.~B {\bf 20}, 12, 4981 (1979).}
\bibitem{tf}{L.H.~Thomas,
   Proc. Cambridge Phil. Soc. {\bf 23}, 5, 542–548 (1927);
   E.~Fermi, 
   Rend. Accad. Naz. Lincei. {\bf 6}, 602–607 (1927).}
\bibitem{Liberman1990}{D.A.~Liberman and B.I.~Bennett, Phys. Rev.~B {\bf 42}, 2475 (1990).}
\bibitem{Swift2018}{D.C.~Swift, T.~Lockard, M.~Bethkenhagen, R.G.~Kraus,
   L.X.~Benedict, P.~Sterne, M.~Bethkenhagen, S.~Hamel, and B.I.~Bennett,
   submitted and {\tt arXiv:1903.00163} (2018).}
\bibitem{Waldram1985}{J.R.~Waldram, ``The Theory of Thermodynamics'', Cambridge (1985).}
\bibitem{simplecv}{For example,
   D.J.~Steinberg,
   Lawrence Livermore National Laboratory report UCRL-MA-106439 change 1 (1996).}
\bibitem{debye}{P.~Debye, 
   Ann. Phys. {\bf 39}, 4, 789–839 (1912).}
\bibitem{Swift2001}{D.C.~Swift, G.J.~Ackland, A.~Hauer, and G.A.~Kyrala,
   Phys. Rev.~B {\bf 64}, 214107, (2001).}
\bibitem{multidebye}{A.A.~Correa, L.X.~Benedict, D.A.~Young, E.~Schwegler, and S.A.~Bonev, Phys. Rev.~B {\bf 78}, 024101 (2008).}
\bibitem{Whitley2015}{H.D.~Whitley, D.M.~Sanchez, S.~Hamel, A.A.~Correa, and L.X.~Benedict,
    Contrib. Plasma Phys. {\bf 55}, 5, pp.~390-398 (2015).}
\bibitem{vasp}{G.~Kresse and J.~Furthm\"uller,
   Phys. Rev. B {\bf 54}, 11169 (1996).}
\bibitem{paw}{P.E.~Bl\"ochl, Phys. Rev. B {\bf 50}, 17953 (1994).}
\bibitem{Baldereschi1972}{A.~Baldereschi, Phys. Rev. B {\bf 7}, 5212 (1972).}
\bibitem{Hohenberg1964}{P.~Hohenberg and W.~Kohn,
   Phys. Rev. B {\bf 136}, 3B, pp~864--871 (1964).}
\bibitem{Kohn1965}{W.~Kohn and L.J.~Sham,
   Phys Rev {\bf 140}, 4A, pp~1133--1138 (1965).}
\bibitem{Perdew1992}{J.~Perdew, Phys Rev {\bf B46}, 6671 (1992).}
\bibitem{Perdew1994}{J.~Perdew, Phys Rev {\bf B50}, 4954 (1994).}
\bibitem{Perdew1996}{J.P.~Perdew, K.~Burke, and M.~Ernzerhof,
   Phys. Rev. Lett. {\bf 77}, 3865 (1996).}
\bibitem{Hoover1996}{W.G.~Hoover and B.L.~Holian,
   Phys. Lett. A {\bf 211}, 5, pp~253–257 (1996).}
\bibitem{Swift2008}{D.C.~Swift, J.~Appl. Phys. {\bf 104}, 073536 (2008).}
\bibitem{Swift2018a}{D.C.~Swift and M.~Millot, in preparation.}
\bibitem{Crockett2006}{S.~Crockett, documentation for {\sc sesame} EOS 7834,
   Los Alamos National Laboratory (2006).}
\bibitem{shockexpts}{M.N.~Pavlovskii, Sov. Phys. Solid State {\bf 13}, 741 (1971);
   K.~Kondo and T.J.~Ahrens, Geophys. Res. Lett. {\bf 10}, 281 (1983);
   D.K.~Bradley et al, Phys. Rev. Lett. {\bf 93}, 19, 195506 (2004);
   H.~Nagao et al., Phys. Plasmas {\bf 13}, 052705 (2006);
   S.~Brygoo et al., Nat. Mater. {\bf 6}, 274 (2007);
   D.G.~Hicks et al, Phys. Rev. B {\bf 78}, 174102 (2008);
   R.S.~McWilliams et al, Phys. Rev. B {\bf 81}, 014111 (2010);
   M.~Knudson et al, Science {\bf 322}, 1822 (2008);
   M.C.~Gregor et al, Phys. Rev. B {\bf 95}, 144114 (2017).}
\end{thebibliography}
\end{document}